\documentstyle[twoside,fleqn,espcrc2]{article}

\newcommand{\twooneplaq}{\setlength{\unitlength}{.5cm}
   \raisebox{-.2cm}{
   \begin{picture}(2.2,1.2)(-1.1,-.6)
   \put(-1,-.5){\line(1,0){2}}
   \put(-1,.5){\line(1,0){2}}
   \put(-1,-.5){\line(0,1){1}}
   \put(1,-.5){\line(0,1){1}}
   \put(-1,-0.5){\circle*{.2}}
   \put(-1.5,-0.9){$x$}
   \put(-0.05,-1.2){$\mu$}
   \put(-1.5,-0.05){$\nu$}
   \end{picture}}}
\newcommand{\ltwooneplaq}{\setlength{\unitlength}{.5cm}
   \raisebox{-.2cm}{
   \begin{picture}(2.2,1.2)(-1.1,-.6)
   \put(-.5,-1){\line(1,0){1}}
   \put(-.5,1){\line(1,0){1}}
   \put(-.5,-1){\line(0,1){2}}
   \put(.5,-1){\line(0,1){2}}
   \put(-0.5,-1){\circle*{.2}}
   \put(-0.9,-1.5){$x$}
   \put(-1.2,-0.05){$\nu$}
   \put(-0.05,-1.5){$\mu$}
   \end{picture}}}

\newcommand{\AmS}{{\protect\the\textfont2
  A\kern-.1667em\lower.5ex\hbox{M}\kern-.125emS}}
\newcounter{arabiclistc}

\def\sqr#1#2#3{{\vcenter{\hrule height.#2pt
      \hbox{\vrule width.#2pt height#1pt \kern#3pt
         \vrule width.#2pt}
      \hrule height.#2pt}}}

\title{Improved cooling algorithm for gauge theories}

\author{Philippe  de  Forcrand\address{IPS, ETH-Z\"urich, CH-8092 Z\"urich,
Switzerland}%
, Margarita  Garc{\'\i}a  P\'erez\address{Instituut Lorentz,
Rijksuniversiteit Leiden, PO Box 9506, NL-2300 RA Leiden, Nederland}%
\thanks{
 MEC and FOM support is thankfully acknowledged. }
and Ion-Olimpiu  Stamatescu\address{FEST, Schmeilweg 5, D-69118 Heidelberg,
Germany
\\  and \\
Inst. Theor. Physik,  Univ. Heidelberg, D-69120 Heidelberg, Germany
}}

\begin{document}

\begin{abstract}
We propose and study a ``gold-washing" - type of algorithm which
smooths out the short range fluctuations but leaves invariant
instantons above a certain size. The algorithm needs no monitoring
or calibration.
\end{abstract}

\maketitle

\section{THE COOLING PROBLEM}

As a method to eliminate UV noise and permit studying
topological excitations cooling must fulfill certain requirements:\par
\noindent - smooth out the short range fluctuations,
including ``dislocations"\par
\noindent - preserve the structure at the physical scales,
including size and location of instantons\par
\noindent - ensure stability of the cooled structures \par
\noindent - need no monitoring or any engineering
which could make it configuration dependent or
would only slow it down without ensuring  stability.\par

Since Wilson action
has no stable instantons (they shrink and decay
under cooling) one usually attempts to calibrate or engineer
the cooling procedure based on it such as to obtain metastability,
after the noise has been reduced but before the
instantons shrank to zero. This, however, makes
cooling more an art than a method and also does not answer
the question of preserving the physical scales of the original configuration.
A general tree level analysis of various actions \cite{MP} distinguishes
between ``under - improved" (e.g., Wilson) and ``over -
improved" ones; under the former instantons shrink,
under the latter instantons beyond a certain size grow \cite{MP,PK}.
We here consider the problem of ``improved" cooling algorithms
which ensure a high degree of scale invariance for the instantons
beyond the short range scale. They should
be useful not only for producing good susceptibility
data, but also for studying various conjectured effects of
instantons for the spectrum or the chiral transition by  providing
smooth configurations which preserve the large scale structure
of the original (hot) ones. This is why
we called this approach ``gold washing".

The question of a good cooling algorithm is of course related to that
of an action possessing scale invariant classical solutions.
``Fix point perfect actions", expected to have generally good
scaling properties
have been constructed
in terms of many loops and
higher representations \cite{Ha}.
One still needs, however, to test their usefulness for cooling algorithms
in gauge theories.

Our approach here has been different. Since we are primarily interested
in instantons our construction refers directly to them. We
start with tree level improvement and
tune the action to obtain stable instanton solutions which are
practically scale invariant
beyond some small-size threshold; the latter should be
such that only physically relevant
instantons are preserved while short distance topological
fluctuations (``dislocations") are eliminated in the cooling.
It turns out that the improved cooling
algorithm which will be presented below fulfills most of the requirements
stated above.
The instantons stabilized by it are well fitted by the continuum ansatz
for volume integrated quantities (we use twisted b.c. in ``time" but the
same should hold if the time extension of the lattice is large enough).
They have integer topological charge within   ${\cal O}( 0.1{\%})$.
Higher charge configurations are well behaved under this algorithm,
however instanton-antiinstanton (I-A) pairs annihilate of course
after a certain
number of steps, due to their interaction. The study of pairs necessitates
therefore further developments. Although we concentrate here on the
cooling problem, the good scaling properties of the actions we study make
them useful also as improved Monte Carlo actions. While our analysis here
refers explicitely to $SU(2)$, most relations are valid for $SU(N)$ generally.

\section{IMPROVED COOLING}

One can take various loops in the action and tune their couplings
to improve the approach to continuum. For simplicity we work
only with fundamental, planar loops of size $m \times n$:

\begin{eqnarray}
S_{m,n}\hspace{-.3cm} &=& \hspace{-.2cm}\sum_{x,\mu,\nu} {\rm Tr}
\left( 1 -\frac{1}{2}\
 ( \   \twooneplaq + \hspace{-.1cm}\ltwooneplaq\ \hspace{-.1cm}
) \right) \\
S\hspace{-.3cm} &=& \hspace{-.2cm}  \sum_{i=1}^5 c_i \
{1 \over {m_i^2 n_i^2}}
\ S_{m_i,n_i}
\end{eqnarray}

\noindent Here  $(m_i,n_i) = (1,1),(2,2),(1,2),(1,3),(3,3)$
for $i=1,\ldots , 5$. The Gibbs factor is ${\rm exp}\ (-{1 \over
{g^2}} S)$.  The choice
\begin{equation}
S(\epsilon ):\ \ c_1 = (4-\epsilon)/3,\ c_2 = (\epsilon -1)/3
\end{equation}
$c_{3,4,5} = 0$, leads to "under-improved" actions for $\epsilon > 0$
(instantons shrink, as for the Wilson action $\epsilon = 1$). For
$\epsilon < 0$ the actions are "over-improved" (instantons grow)
and for $\epsilon = 0$ there are no ${\cal O}(a^2)$ corrections \cite{MP}.
Using the results in  \cite{MP} one can construct a one parameter
set of actions that  have no ${\cal O}(a^2)$ and ${\cal O}(a^4)$ corrections
\begin{eqnarray}
c_1 &=&\hspace{-0.2cm}(19- 55\  c_5)/9,\ \  c_2 =  (1- 64\  c_5)/9 \nonumber\\
c_3 &=&\hspace{-0.2cm}(-64+ 640\  c_5)/45,\ \  c_4 = 1/5 - 2\  c_5
\end{eqnarray}
The simplest ones are $S(4Li)\ (c_5\hspace{-0.1cm}=\hspace{-0.1cm}0)$
and $S(3Li)$$(c_5\hspace{-0.1cm} =\hspace{-0.1cm}1/10)$.
Since for the configurations we have studied numerically
the former tends to under-improve while the latter over-improves,
we chose for our simulations a combination of both, $S(5Li)$, with
$c_5=1/20$. This action showed already good properties and we
did not try to optimize it further.
The cooling algorithm exactly minimizes the above local action at each
step and involves no further calibration or engineering.
\begin{figure}[htb]
\vspace{4.4cm}
\includegraphics{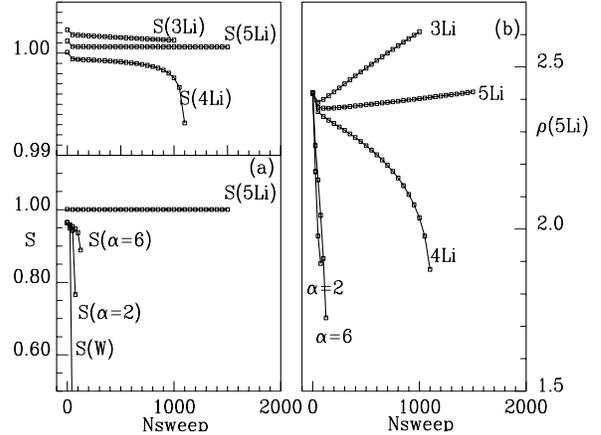}
\caption{Action (a) and size (b) {\it vs} sweep number for various algorithms.
S(W) is Wilson action.}
\label{fig 1}
\end{figure}

\begin{figure}[htb]
\vspace{5.34cm}
\includegraphics{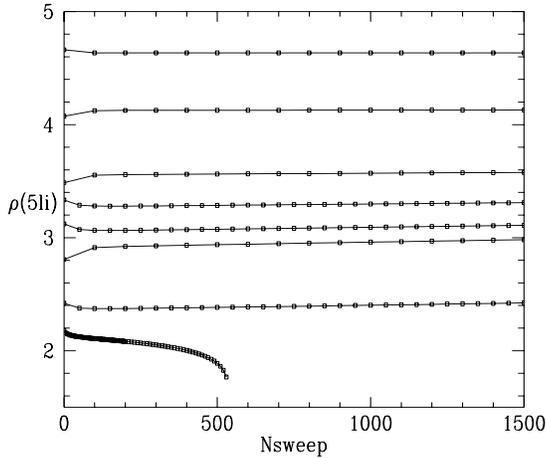}
\caption{Instanton size under $5Li$ improved cooling starting from various
configurations.}
\label{fig 2}
\end{figure}

We first check the behaviour of various instantons under the
improved cooling taking as starting
configurations those obtained by cooling with $S(\epsilon )$ with various
$\epsilon$. All results are obtained on $12^4$ lattices and for gauge
group $SU(2)$. Since we wanted to disentangle small distance from finite
lattice size effects, we use throughout twisted p.b.c. with $k=(1,1,1)$
twist, to ensure that instanton solutions exist on finite lattices
\cite{BP,MP}. Fig. 1 compares the cooling behaviour of
some algorithms ($S(\alpha)$ denote the underrelaxed
algorithms of \cite{MS}): (a) the action (notice the
different scales) and (b) the size $\rho_{peak} = 6\pi^2N_s / S(t)|_{max}$
for the same instanton, as function of the cooling step ($N_s=12$ here).
Fig. 2 shows the behaviour of instantons of various initial sizes
under the $5Li$-cooling. As it can be seen, they
remain practically unchanged over any practicable
number of cooling sweeps and for  $\rho > \rho_0 \simeq 2.3$
(in units of a), which seems to represent a stability
threshold for this  cooling algorithm.
The size of the original instanton has been
varied by applying $S(\epsilon)$ cooling, therefore these instantons do not
correspond to minima of $S(5Li)$. Submitting such configurations
to $5Li$ cooling first readapts them to the new equations of motion,
implying the small changes which can be observed in the first $\simeq 50$
sweeps of Fig. 2.  Fig. 3 shows the effect of $5Li$ cooling on instantons
of different sizes taken from various stages of Wilson cooling of 3 different
starting configurations (the upper curves give
the values of $S(5Li)$ for these Wilson cooled instantons).
 We see how the $S(5Li)_{min}$ emerges as an envelope.
We observe again the stability threshold around 2.3 and the extent
of  configuration dependence at small $\rho$. The triangles are data
from Monte Carlo instantons ($\beta=2.5,\ 12^4$ lattice) cooled with 300
sweeps of $5Li$ cooling.
\begin{figure}[htb]
\vspace{5.34cm}
\includegraphics{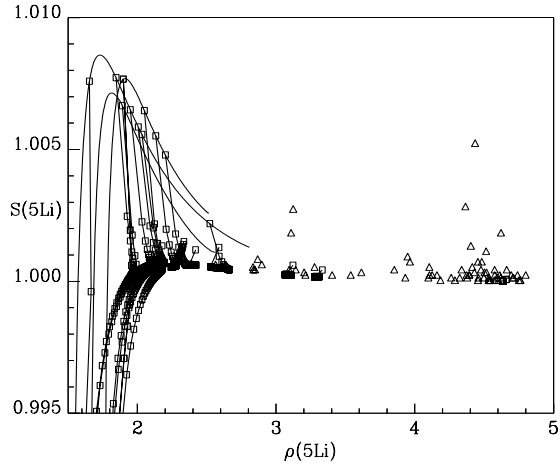}
\caption{Instanton evolution (action and size) during $5Li$ cooling.}
\label{fig 3}
\end{figure}

To describe the instantons we use the volume
integrated action and charge densities (``profiles"):
\begin{eqnarray}
S(t) &=& N_s\sum_{x,y,z} S(x,y,z,t) \\
Q(t) &=& 8\pi^2N_s\sum_{x,y,z} Q(x,y,z,t)
\end{eqnarray}

\noindent
to be fitted by the continuum ansatz with periodicity satellites, e.g.
for the charge:
\begin{eqnarray}
Q(t)\hspace{-.2cm}  &=& \hspace{-.2cm}\pm 6\pi^2N_s\rho^4
[f(t) +f(t+N_t) + f(t-N_t)] \nonumber\\
& & f(t) = [(t-t_0)^2+\rho^2]^{-5/2}
\end{eqnarray}

\noindent  Here $t_0$ defines the location and $\rho$ the width,
``$\rho_{profile}$". The latter agrees rather well with $\rho_{peak}$ as
long as the total charge is  $\pm 1$.
This is ensured at not too small $\rho$, i.e. for stable \mbox{instantons}.

Since Wilson action produces short range fluctuations one
can ask whether the effective cut-off at   $\rho_0 \simeq 2.3$
 ensures the absence of unphysical fluctuations in cooled MC configurations.
Following an argument of Pugh and Teper \cite{PT} we write the
contribution of small instantons as ($N=2$ here)
\begin{eqnarray}
\left[\rho_0(a)a\right]^{-4}\hbox{e}^{-\beta S_W(\rho_0)} \\
a(\beta) \simeq \Lambda^{-1} \hbox{e}^{-{{\beta} \over {4N\beta_1}}};\ \
\beta_1 = {{11N} \over {48\pi^2}} \nonumber
\end{eqnarray}

\noindent
We can calculate $S_W(\rho_0)$ by cooling large
instantons with the Wilson action (they shrink).
One can then find $\rho_0 (a(\beta ))$ from the condition
that the two factors in Eq.(8) compensate each other such that the
contribution from instantons larger than $\rho_0$ stays finite in continuum.
We obtain a very flat dependence: $\rho_0 = 1.9276, 1.9280,  1.9284$ for
$\beta = 3,  6,  \infty$, respectively.
Hence a threshold $\rho_0$ of about 2 and independent on $\beta$
seems satisfactory.

\begin{figure}[htb]
\vspace{5.3cm}
\includegraphics{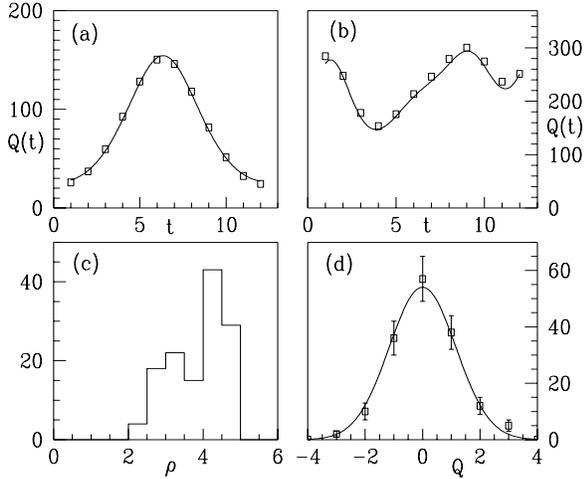}
\caption{MC analysis:  typical charge 1 and 3
configurations and size and charge distribution.}
\label{fig4}
\end{figure}

\begin{figure}[htb]
\vspace{5.3cm}
\includegraphics{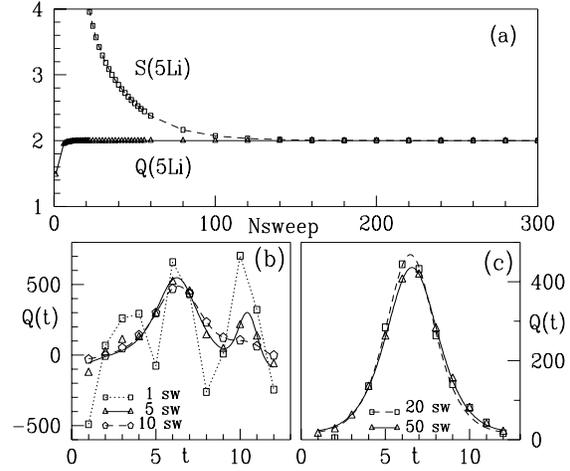}
\caption{Early evolution of a MC conf.. The line for 1 sw
is an eye-guide, the rest fits based on (7).}
\label{fig 5}
\end{figure}

\section{MONTE CARLO ANALYSIS}

To illustrate the capability of  improved cooling we apply
$5Li$ cooling to a Monte Carlo simulation at $\beta=2.5$ on a $12^4$
lattice, $k=(1,1,1)$ twist.
We use  160 configurations generated by heat bath with Wilson action and
taken 250 sweeps apart (after 20000 thermalization sweeps).
In Fig. 4 we show results from 300 cooling sweeps. The stable situation
typically sets in between 50 and 100 sweeps.
Instantons very near below the threshold $\rho_0 \simeq 2.3$  may need
more sweeps to disappear - see Fig. 2; this introduces a systematic error
in the number of stable instantons in this region, which
vanishes with increasing number of sweeps.
In Fig. 4a,b we show two typical configurations, the first  with one wide
($\rho_{profile}= 4.66$) instanton  and the second with  total charge
3 showing 3 instantons of widths 2.80, 4.47 and 3.37 located at
t=1.22, 6.22 and 9.29, together with the fit Eq.(7).
In Fig. 4c we show the size and in 4d the charge distributions.
The latter is representative for physical instantons, i.e.
beyond the threshold $\rho_0$.
A gaussian fit ${\rm exp}(-bQ^2)$ gives $b=0.34$ at cooling sweep 20 and
$b=0.37$ beyond 100. The charge stabilizes
to within $1\%$ an integer between 20 and 100 sweeps, therefore
susceptibility and charge distribution can be estimated quite early.
The topological susceptibility extracted from  data at 300 sweeps is
$Q^2/12^4 = 6.8(9)\  10^{-5}$. The size distribution, however,
represents only the stable situation and does not take into account
I-A pairs which would have annihilated before.
For illustration we show in Fig. 5 the charge and action
history (a) and the evolution of a typical configuration under the
$5Li$ cooling (b,c). The initial structure seems to have a double instanton
and 2 I-A pairs. An I-A pair can still be seen with the continuum ansatz
Eq. (7) at sweep 5 (sizes 1.38 and 2.95) and at sweep 10
(sizes 2.74 and 3.36), together with the double instanton (size 2.5 at
sweep 5 and 2.86 at 10) - see Fig. 5b. After sweep 20 (Fig. 5c) only the
latter remains (size: 3.07 at 20, 3.29 at 50, 3.52 at 300 sweeps).
Hence this annihilation may result in a depletion of the lower part of
the size distribution.  The question of describing complex configurations
is a separate problem currently under study.

We are very indebted to  Pierre van Baal and Jeroen Snippe for discussions
and suggestions. MGP and IOS profitted very much from the participation
to the Benasque Center of Physics which gave the opportunity to
work concentrated on this subject.
Use of the Fujitsu VP600 of the  University of  Karlsruhe
and  a NCF grant for use of the Cray C98 are thankfully
acknowledged.

\end{document}